\documentclass[useAMS]{mn2e}

\input{epsf}

\title{Fitting optical source counts with an infrared-defined model: insights into galaxy evolution}
\author[A. J. King and M. Rowan-Robinson]
       {A. J. King\thanks{E-mail: alex.king@ic.ac.uk} and M. Rowan-Robinson\\
        Astrophysics Group, Blackett Laboratory, Imperial College of Science, Technology and Medicine, Prince Consort Road, London SW7 2BW, UK}
\date{Accepted 0000 January 00.
      Received 0000 January 00;
      in original form 2002 March 20}

\pagerange{\pageref{firstpage}--\pageref{lastpage}}
\pubyear{2002}

\begin{document}

\maketitle

\label{firstpage}

\begin{abstract}

In order to improve the fit to the optical counts, the multi-wavelength source count model of Rowan-Robinson (2001) is improved in two ways. First, density evolution is incorporated to correspond to the mergers in hierarchical models of galaxy formation. Secondly, the evolution of dust opacity is taken into account. The luminosity and density evolution parameters and the characteristic opacity of the local universe are allowed to vary freely and the best overall fit to the {\it B} band is determined. This parameter set is then translated to other wavebands, showing significant improvement of goodness of fit in the optical, UV and near-IR bands, whilst retaining the goodness of fit in the far-IR and submillimetre bands of the original model. A significant fraction of the improvement of fit is shown to be due to the introduction of optical depth history. The mean value of $A_{V}$ in the local universe for the best-fitting model is 0.4. The goodness of fit to available star formation history data is not significantly altered and the form of the evolution of the luminosity function shown to be consistent with a semi-analytic, forward evolution model.
\end{abstract}

\begin{keywords}
cosmology: observations -- dust, extinction -- galaxies: evolution -- galaxies: ISM -- infrared: ISM -- stars: formation 
\end{keywords}


\section{Introduction}
\label{sec:intro}

The cosmic history of star formation has been extensively investigated in recent times. There is a general consensus that the star formation rate (SFR) increases with look-back time to peak somewhere at $1 < z < 2$, then levels off, with a possible decline at higher redshifts. However, the issue is far from closed. In particular, the location of the peak and the behaviour of the SFR at high redshifts are still a matter of extensive debate.

The nature of the star formation history (SFH) can be determined using a number of different observational techniques. Perhaps the most direct method is to make use of observations in the rest-frame near-ultraviolet (UV), as this part of the spectra of galaxies with no active galactic nuclei (AGN) are dominated by emission from short-lived, massive stars (e.g. Lilly et al. 1996; Madau et al. 1996; Connolly et al. 1997; Madau, Pozzetti \& Dickinson 1998; Treyer et al. 1998; Cowie, Songaila \& Barger 1999). It is also possible to observe the luminosities of nebular emission lines (especially H$\alpha$), which are due to reprocessed ionizing radiation from the young, massive stars (e.g. Hogg et al. 1998; Tresse \& Maddox 1998; Glazebrook et al. 1999; Hopkins, Connolly \& Szalay 2000; Teplitz et al. 2000; Pettini et al. 2001). Neither of these methods, however, account for the effects of dust extinction.

The importance of extinction has been shown by numerous studies that show the disagreement of different SFH diagnostics, which are affected by extinction in different ways (e.g. Mushotzky \& Loewenstein 1997; Rowan-Robinson et al. 1997; Cram et al. 1998; Flores et al. 1999; Adelberger \& Steidel 2000; Sullivan et al. 2001) and from the large fraction of energy in the cosmic far-infrared and submillimetre backgrounds (Puget et al. 1996). Estimates of the extinction correction factor are around 2--10 (e.g. Meurer et al. 1997; Dickinson 1998; Pettini et al. 1998; Glazebrook et al. 1999; Steidel et al. 1999; Hopkins et al. 2000; Teplitz et al. 2000). It is, however, possible that some star formation could be totally hidden in heavily obscured objects, which would not be accounted for at all by a correction to the UV/optical estimates (Franceschini et al. 1997; Guiderdoni et al. 1998; Elbaz et al. 1999; but see Adelberger \& Steidel 2000). Furthermore, it has recently been shown that the far-infrared (FIR) to UV flux ratio breaks down for some star-forming galaxies (Bell \& Kennicutt 2001), ultraluminous infrared galaxies (ULIRGs; Meurer et al. 2000) and Lyman Break Galaxies (e.g. Baker et al. 2001), which makes it difficult to determine the SFR by simply applying an extinction correction.

To avoid these problems, it is necessary to look at the radiation re-emitted from dust heated by the young stellar population, which will also reveal any obscured star-forming population. Since thermal dust emission peaks in the rest-frame far-infrared (IR), this is done with submillimetre and far-IR surveys (e.g. Rowan-Robinson et al. 1997; Hughes et al. 1998; Blain et al. 1999b). The main drawback with this method is the uncertainty of the contributions of cumulative heating of the dust by the old stellar population and AGNs (Blain et al. 1999c; Brusa, Comastri \& Vignali 2001).

A very different way of approaching the issue of how to determine the SFH is to use source count models. Broadly speaking, these fall into one of two categories: the so-called `forwards' and `backwards' models (Lonsdale 1996). The forwards method statistically follows the growth of dark matter haloes forwards in time by accretion and mergers (e.g. Kauffmann \& White 1993; Cole et al. 1994; Blain et al. 1999c; Granato et al. 2000; Somerville, Primack \& Faber 2001) then assumes a semi-analytic galaxy formation paradigm and evolutionary framework. In the backwards method, by contrast, a local luminosity function is evolved backwards in time, assuming some form of evolution, such as a change in SFR (e.g. Pearson \& Rowan-Robinson 1996; Eales \& Edmunds 1997; Xu et al. 1998; Takeuchi et al. 1999; Pearson 2001; Rowan-Robinson 2001; Xu et al. 2001). 

In Rowan-Robinson (2001; hereafter RR01), a source count model using a parametrized approach to the SFH was presented. This method models the SFH as a combined power-law and exponential, the relative strengths of which are allowed to vary freely. This is capable of reproducing most physically realistic forms of the SFH and allows a large set of possible histories to be investigated. Source count predictions are then compared to observations, allowing the range of SFH parameters to be limited. This method is similar in some ways to those of Guiderdoni et al. (1998) and Blain et al. (1999b).

The parametrized approach of RR01 provided good fits to data in the submillimetre and far- and mid-IR regions of the spectrum ($10\ \umu \textrm{m}< \lambda < 1250\ \umu$m). However, the best-fitting model shows a deficit of faint ($B > 24, K > 20$) optical and near-IR sources compared to observations. Also, the number of high redshift sources are over-predicted. RR01 proposes that these problems could be caused by density evolution, or by a steepening of the faint-end luminosity function at $z > 1$. 

In this work, an attempt to solve the faint optical count issue by allowing a variation of optical depth with redshift using the Calzetti \& Heckman (1999) evolution model and by including density evolution is presented. The objective is to produce an effective, multi-wavelength source count model, based on that detailed in RR01. 

The layout of this paper is as follows. In \S\ref{sec:sc}, a brief explanation of the RR01 parametrized source count model is presented. Then, in \S\ref{sec:optdep}, the optical depth history correction is detailed. Section \ref{sec:fitoptcounts} discusses the method of fitting to the optical counts and presents the best-fitting model. Section \ref{sec:predcounts} shows the predicted counts in other wavebands and the background spectrum, with comparisons to available data. In \S\ref{sec:discussion}, the nature of the fits to source count data, the nature of the optical depth history, the implications for the SFH and galaxy evolution and comparisons to a semi-analytic model are discussed. Finally, the conclusions are presented in \S\ref{sec:conclusions}.


\section{Parametrized Source Count Model}
\label{sec:sc}

A backwards source count model can be constructed using four basic ingredients: a cosmology, a luminosity function, a form of evolution (luminosity and/or density) and a spectral energy distribution (SED). In this work, the model of RR01 is used as a base: a more detailed description of the model can be found in that paper. For computational speed, an Einstein--de Sitter ($\Omega_{0} = 1$, $\Lambda = 0$) Universe with $H_{0} = 100$ kms$^{-1}$Mpc$^{-1}$ is examined first. The results of this are used to limit the free parameter space of the Lambda cosmology ($\Omega_{0} = 0.3$, $\Lambda = 0.7$).


\subsection{Luminosity Function}
\label{subsec:lumfunc}

The luminosity function used is determined from the {\it IRAS} PSCz sample (Saunders et al. 2000). This is then fitted with the empirical form of Saunders et al. (1990) for a single, composite population: 

\[
 \eta(L) = \frac{d\Phi}{d\lg L} =
 C_{*} \left( \frac{L}{L_{*}} \right)^{1-\alpha} \exp \left[ - \frac{1}{2\sigma^{2}} \lg^{2} \left( 1+\frac{L}{L_{*}} \right) \right], 
\]
where $C_{*}, \alpha$ and $\sigma$ are constants. This behaves as a power law for $L \ll L_{*}$ and as a Gaussian in log L for $L \gg L_{*}$, $L_{*}$ being the `knee' of the curve. For $\alpha$ and $C_{*}$, fixed values of $\alpha=1.09$ and $C_{*} = 0.027(H_{0}/100)^{3}$ have been assumed. The other parameters take into account the change of the 60-$\umu$m luminosity function as the rate of evolution is varied (see \S\ref{subsec:evolution}).


\subsection{Parametrized form of Evolution}
\label{subsec:evolution}

Changes in the SFR within a galaxy will affect its luminosity. RR01 assumes that the changes of SFR are the dominant factor in luminosity evolution, and parametrizes this evolution to allow a wide range of model star formation histories to be explored. The form of the star formation rate, $\dot{\phi_{*}}(z)$ (in units of M$_{\odot}$ yr$^{-1}$ Mpc$^{-3}$), adopted to allow most physically realistic, single population SFH scenarios to be reproduced is:

\[
 \label{eqn:sfh}
 \frac{\dot{\phi_{*}}(t)}{\dot{\phi_{*}}(t_{0})} = \left[\exp\:Q\left(1-\frac{t}{t_{0}}\right)\right]\left(\frac{t}{t_{0}}\right)^{P},
\]
where $P$ and $Q$ are positive variable parameters (see fig. 1 of RR01 for examples of this form). In an Einstein-de Sitter (E-dS) Universe, this becomes:

\begin{equation}
 \frac{\dot{\phi_{*}}(z)}{\dot{\phi_{*}}(z_{0})} = \left\{\exp\:Q\left[1-(1+z)^{-3/2}\right]\right\}\left(1+z\right)^{-3P/2}.
\end{equation}

The physical meanings of the SFH parameters are as follows. $P$ measures how steeply the process of buildup of rate of star formation due to mergers of fragments, which are incorporated into galaxies, occurs with time. $Q$ describes an exponential decay in the star formation rate, on a timescale $t_{0}/Q$, which is a product of the process of exhaustion of available gas for star formation -- this being the result of a competition between the processes of gas being formed into stars and it being returned to the ISM by winds and supernovae. 

It is assumed, for computational purposes, that $\dot{\phi_{*}}(z) = 0$ for $z > 10$. Although this introduces an unphysical discontinuity in eqn. (\ref{eqn:sfh}), RR01 finds that the effect on the results is negligible.

Now that the luminosity evolution has been selected, the remaining parameters in the luminosity function can be approximated in the E-dS case by least-squares fitting the Saunders et al. (1990) form to the Saunders et al. (2000) data set as follows:

\[
 \sigma = 0.744 - 0.033W,
\]

\[
 \lg{L_{*}} = 8.50 + 0.05W - 2 \lg{(H_{0}/100)},
\]
where 

\[
 W = Q - 1.125P.
\]

In a hierarchical galaxy-formation scenario, it is expected that a lower total density of galaxies will be seen at low redshifts as a result of merger events. In addition to the luminosity evolution, this density evolution also needs to be considered. This can be parametrized simply as:

\begin{equation}
 \label{eqn:density}
 \rho(z) = \rho(0)(1+z)^{n}.
\end{equation}
In this case, the evolution parameter, $W$, becomes

\[
 W = Q - 1.125(P-0.5n).
\]

Many source count models consider only pure luminosity evolution or pure density evolution, but here both forms will be considered simultaneously.

In the Lambda cosmology with density evolution, these parameters become:

\[
 \sigma = 0.783 - 0.030W,
\]

\[
 \lg{L_{*}} = 8.39 + 0.05W - 2 \lg{(H_{0}/100)},
\]
where 

\[
 W = Q - 1.067(P-0.5n).
\]


\subsection{Spectral Energy Distributions}
\label{subsec:seds}

To transform the 60-$\umu$m luminosity function to other wavelengths, some model of the SED must be constructed. RR01 explored a variety of assumptions about the SEDs. The method that was most consistent with available data was to derive the counts separately for each of four components -- starburst (M82-like), `cirrus' (normal, non-active galaxy), ULIRG (Arp 220-like) and AGN -- then sum them.

For the starburst and cirrus components, predictions for infrared SEDs from Efstathiou, Rowan-Robinson \& Siebenmorgen  (2000) are used, and near-IR/optical/UV SEDs are added, corresponding to an Sab galaxy (Yoshii \& Takahara 1988) for the cirrus component and an H$\,$\textsc{ii} galaxy, Tol 1924 -- 416 (Calzetti \& Kinney 1992), for the starburst component. These are selected as `typical' galaxies, rather than as averages. For the cirrus component, the optical SED has been divided into a contribution of young, high-mass stars (dominating at $\lambda < 0.3\ \umu$m) and of old, low-mass stars (dominating at $\lambda > 0.3\ \umu$m). This is obviously a quite approximate treatment, but produces reasonable counts in the {\it K}- and {\it B}-bands. For the AGN component, the dust torus model of Rowan-Robinson (1995) is used (see also Rowan-Robinson \& Efstathiou 1993 and Efstathiou \& Rowan-Robinson 1995). The complete SEDs are shown in fig. 3 of RR01.

The proportions of the four components at 60-$\umu$m as a function of luminosity were chosen to give correct relations in colour-colour diagrams, as detailed in RR01. The relative proportion of the 60-$\umu$m emission due to AGNs is derived from the 12-$\umu$m luminosity function of Rush, Malkan \& Spinoglio (1993).


\section{Optical depth history}
\label{sec:optdep}

The models of RR01 fit the available source count data well from mid-infrared to submillimetre wavelengths. However, in the {\it K}-band, the number of faint ($K > 20$) sources is under-predicted (see fig. 18 of RR01; also plotted on Fig. \ref{fig:KNS} of this work) by the preferred $P = 1.2, Q = 5.4$ model. This problem is even more pronounced in the {\it B}-band, where the faint ($B > 24$) counts are significantly under-predicted (see fig. 19 of RR01; also plotted on Fig. \ref{fig:BNS} of this work) by the same model.

This faint blue excess could be caused by a number of factors, such as uncertainties in the faint end of the local luminosity function, some form of evolution or a missing population of objects (see review by Ellis 1997). One factor that is known to occur, and could cause a faint blue excess, is the variation of optical depth with redshift (Calzetti 1999; Pei, Fall \& Hauser 1999). It is probable that the opacity changes differently for the various types of sources, but we have used a `characteristic' opacity variation, in the manner of Calzetti 1999 and Pei et al. 1999, as this reduces the number of free parameters.

Star formation produces stars with a range of masses, some of which will be sufficiently massive to explode as supernova very quickly ($\sim 10^{6}$ years; Leitherer \& Heckman 1995), injecting heavy metals into the interstellar medium and producing large quantities of dust on short time scales ($\sim 100$ -- $200$ Myr, e.g. Jones et al. 1994; Dwek 1998). The global star formation rate varies with redshift, so the co-moving density of metals, $\Omega_{m}$, and hence the dust density, will also vary. This variation takes the form of an increase with redshift, peaking somewhere between $z \sim 1$ and $z \sim 3$, then declining (see fig. 8, bottom panel of Pei et al. 1999). This form is the result of a competition between the release of metals into the interstellar medium (ISM) and the metals being taken-up from the ISM as the gas is locked into stars.

As the opacity of the ISM  is determined by the combination of the dust column density (dependent on $\Omega_{m}$ and the gas column density), the extinction curve and the dust distribution, it is possible to model the opacity change with redshift. This new factor can be included in the source count model of RR01 by altering the luminosity distance accordingly.

Initially, the obscuration model of Pei \& Fall (1995) was investigated, using the dust model of Rowan-Robinson (1992) and the colour excess model of Pei et al. (1999), assuming the local mean fraction of starlight absorbed by dust, $A_{V_{0}} = 0.4$. Here, the obscuration of the flux, $F$, is parametrized as:

\[
 F_{obs}(\lambda) = F_{0}(\lambda) 10^{(-0.4 \times 0.4 (E_{S}(B-V)) k(\lambda)/3.1)},
\]
where $E_{S}(B-V)$ is the colour excess and $k(\lambda)$ is the dust obscuration factor.

However, this model does not solve the faint excess problem, as the opacity drops at $z > 1$, and we are trying to reduce the number of faint (hence mostly high redshift) galaxies. A model is required which predicts that the optical depth does not drop until a higher redshift. In order to meet this criterion, the Calzetti (1999) obscuration model was investigated, using model 3 from Calzetti \& Heckman (1999) for $E(B-V)$, as this does not begin to drop until $z > 3$ (see fig. 9, top panel of Calzetti \& Heckman 1999). The Calzetti (1999) model analytically parametrizes the `net' obscuration as:

\begin{equation}
  \label{eqn:obs}
  F_{obs}(\lambda) = F_{0}(\lambda) 10^{-0.4 E_{S}(B-V) k(\lambda)},
\end{equation}
based on observations of high redshift galaxies, where:

\[
 k(\lambda)= 2.656\left(-2.310 + \frac{1.315}{\lambda}\right) + 4.88
\]
for $0.63\ \umu \textrm{m} \leq \lambda \leq 1.60\ \umu \textrm{m}$, and

\[ 
 k(\lambda) = 2.656 \left( -2.156 + \frac{1.509}{\lambda} - \frac{0.198}{\lambda^{2}} + \frac{0.011}{\lambda^{3}} \right) +4.88
\]
for $0.12\ \umu \textrm{m} \leq \lambda < 0.63\ \umu \textrm{m}$.

Outside this range of $\lambda$, the Rowan-Robinson (1992) dust form is used. This is based on a multigrain model, fitted to the interstellar extinction curve and the far-IR spectrum of dust in our galaxy. A look-up table of values of $A_{\lambda}/E(B-V)$ for different wavelengths is used, interpolating where necessary.

This correction for the optical depth history requires a new parameter, $A_{V_{0}}$, to be introduced into the source count model. This represents the mean {\it V} band extinction in the local ($z=0$) universe and can be varied whilst retaining the dependence on redshift of the Calzetti \& Heckman (1999) model 3 extinction, $g(z)$. It is introduced into the net obscuration (eqn. \ref{eqn:obs}) as follows:

\[
  F_{obs}(\lambda) = F_{0}(\lambda) 10^{-0.4 \times A_{\lambda}(z)},
\]
where 

\[
  A_{\lambda}(z) = A_{V_{0}} \times \left( \frac{A_{\lambda}}{A_{V}} \right) \times g(z).
\]

The effect of the optical depth variation introduced here is wavelength-dependent. The magnitude of the effect is smaller at higher wavelengths. It was found that at wavelengths of approximately 5-$\umu$m and higher the effect is negligible. For computational speed, therefore, the correction was applied only at $\lambda < 10\ \umu$m.


\section{Fitting Optical Counts}
\label{sec:fitoptcounts}

The source count model, with the optical depth history correction, has four free parameters: the characteristic visual extinction in the local universe, $A_{V_{0}}$, the density evolution parameter, $n$, and the two star formation parameters, $P$ and $Q$. To determine if the opacity correction could solve the fitting problem in the {\it B} band, these parameters were varied and the resulting model predictions compared to the available observations.

For a given waveband, there are two types of data available against which the models can be compared: source count data and redshift distribution data. Generally, it is possible to obtain large statistics for source counts much more easily than for redshift distributions (few spectroscopic redshifts are available and photometric redshifts can be unreliable). As the source counts are much more robust, statistically speaking, the procedure adopted was to first use the available redshift data -- a catalogue of photometric and spectroscopic redshifts in the Hubble Deep Field (Fern{\' a}ndez-Soto, Lanzetta \& Yahil 1999) derived from data at 0.3, 0.45, 0.6 and 0.8 $\umu$m (filters F300W, F450W, F606W and F814W, respectively) and in the {\it J, H} and {\it K} bands (1.2, 1.6 and 2.2 $\umu$m, respectively) -- to constrain the parameter space to those sets that were as good as, or provided an improvement over, the $P = 1.2, Q = 5.4$ E-dS model of RR01. This was done by generating a large number of models ($\sim 2500$) with a wide range of different parameters, then using a reduced $\chi^{2}$ method to remove those that have a worse fit than the $P = 1.2, Q = 5.4$ model. 

As the number of sources at high redshift for the available data is small, the bins $z = 4.0$ -- $4.5, 4.5$ -- $5.0$ and $z > 5.0$ are summed up to give one bin, $z > 4.0$. This gives the number of model points, $n_{m} = 12$ and the number of free parameters, $n_{c} = 4$ ($P, Q, n$ and  $A_{V_{0}}$) and, hence, the number of degrees of freedom, $\nu = 8$. Around 200 models had better fits than the $P = 1.2, Q = 5.4$ model.

After a reasonable fit to the redshift distributions had been obtained, the source count data was investigated to find the best-fitting model. The data used are compiled from Anglo-Australian Telescope (AAT) observations to a depth of $B \sim 23.5$ (Jones et al. 1991), {\it Hubble Space Telescope} observations of the Hubble Deep Fields (North and South) and William Herschel telescope (WHT) observations to a depth of $B \sim 28.5$ (Metcalfe et al. 2001), and observations made during the commissioning phase of the Sloan Digital Sky Survey to a depth of $B \sim 20$ (Yasuda et al. 2001).

\begin{figure} 
 \epsfysize=7.5cm
 \epsfxsize=8.0cm
 \epsffile{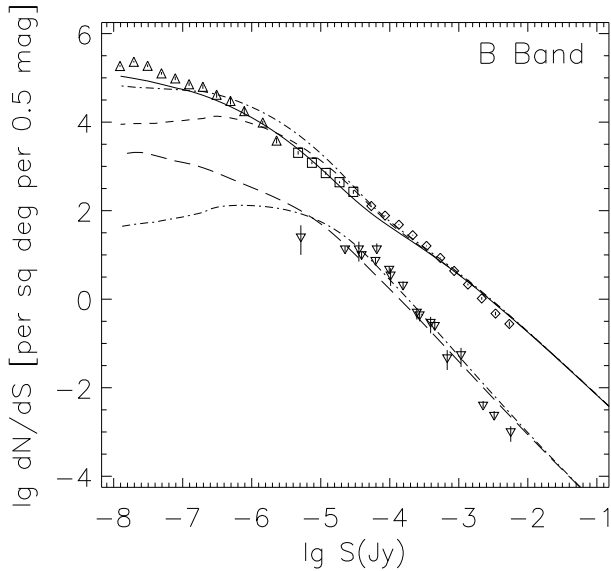}
 \caption{Differential source counts at 0.44-$\umu$m. Data from Jones et al. (1991) [squares], Metcalfe et al. (2001) [triangles] and Yasuda et al. (2001) [diamonds]. Models are the best-fitting E-dS from RR01, $P = 1.2, Q = 5.4$ [short-dashed curve], the best-fitting E-dS from this work, $P = 1.5, Q = 5.4, n = 1.3, A_{V_{0}} = 0.4$ [solid curve] and the comparative model $P = 1.5, Q = 5.4, n = 1.3, A_{V_{0}} = 0.0$ [dotted-dashed curve]. Lower curves are predictions for quasars in the $P = 1.2, Q = 5.4$ [long-dashed curve] and $P = 1.5, Q = 5.4, n = 1.3, A_{V_{0}} = 0.0$ [dotted-dashed curve]. Quasar data from Boyle, Shanks \& Peterson (1988) [inverted triangles]. The goodness of fit is improved in the new model, due to both the changes in evolutionary parameters, and the introduction of the optical depth correction (see text for discussion).}
 \label{fig:BNS}
\end{figure}

\begin{figure}
 \epsfysize=7.5cm
 \epsfxsize=8.0cm
 \epsffile{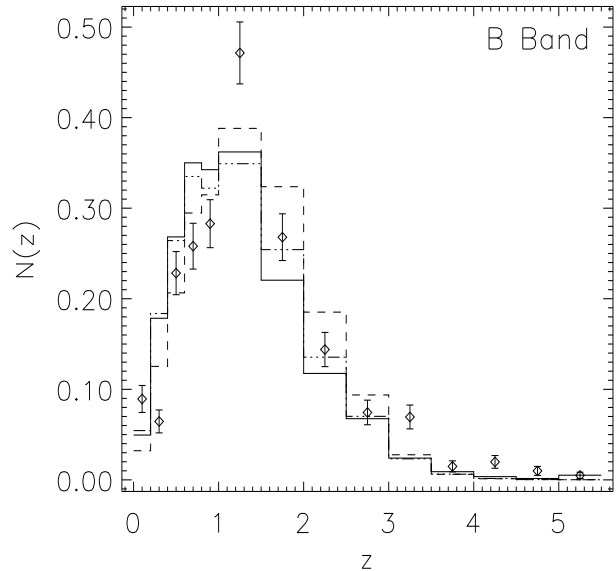}
 \caption{Predicted redshift distribution at 0.44-$\umu$m, flux range $-5.2 > lg S(Jy) > -6.8$ (corresponding to $22 < B < 26$). Models are the best-fitting from RR01, $P = 1.2, Q = 5.4$ [dashed curve], the best-fitting from this work, $P = 1.5, Q = 5.4, n = 1.3, A_{V_{0}} = 0.4$ [solid curve] and the comparative model $P = 1.5, Q = 5.4, n = 1.3, A_{V_{0}} = 0.0$ [dash-triple dot curve]. Data from Fern{\' a}ndez-Soto et al. (1999). Errors are $\sqrt{N}$ Poisson distribution values. The bin shown as $z=5.0$ -- $5.5$ is $z > 5.0$.}
 \label{fig:Bz26}
\end{figure}

For the models that showed an improved fit to the redshift distribution, the $\chi^{2}_{\nu}$ method was again used to find those that had a better fit to the source count data than the $P = 1.2, Q = 5.4$ model. In this case, 4 model parameters are compared to the 28 data points, giving $\nu = 24$. All but six of the parameter sets examined showed a better fit than the $P = 1.2, Q = 5.4$ model. Of these, the best-fitting model was $P = 1.5, Q = 5.4, n=1.3, A_{V_{0}}=0.4$. As can be seen from Figs. \ref{fig:BNS} and \ref{fig:Bz26}, this model gives a reasonable redshift distribution whilst significantly improving the fit to the source count data. Since the luminosity and density evolution parameters have been altered, these figures also show a model with the same evolutionary parameters, but no optical depth correction ($P = 1.5, Q = 5.4, n=1.3, A_{V_{0}}=0.0$). This shows that the improvement of goodness of fit is not due solely to the change in evolutionary parameters, but relies on the new optical depth history.

\begin{figure} 
 \epsfysize=7.5cm
 \epsfxsize=8.0cm
 \epsffile{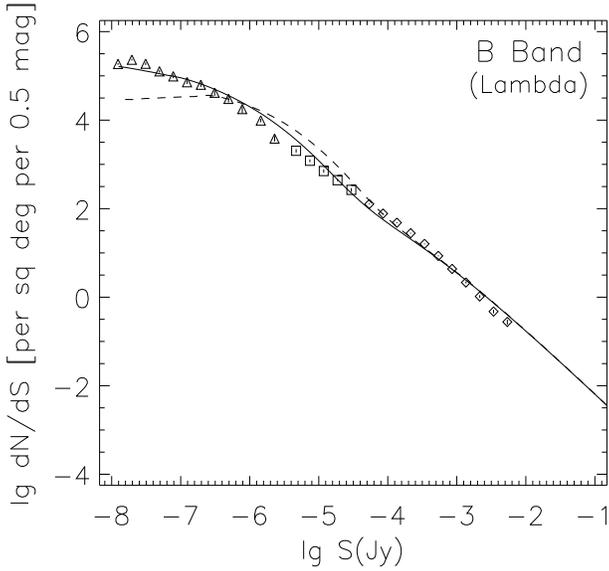}
 \caption{Differential source counts at 0.44-$\umu$m in a $\Lambda$ cosmology. Data from Jones et al. (1991) [squares], Metcalfe et al. (2001) [triangles] and Yasuda et al. (2001) [diamonds]. Models are the best-fitting from RR01, $P = 3.0, Q = 9.0$ [dashed curve] and the best-fitting from this work, $P = 3.4, Q = 9.0, n = 1.1, A_{V_{0}} = 0.4$ [solid curve].}
 \label{fig:BNSL}
\end{figure}

Once this had been done for the E-dS cosmology, the Lambda model was investigated. The best-fitting Lambda model of RR01 had SFH parameters $P = 3.0, Q = 9.0$. As the Lambda model with density evolution is very computationally intensive, a `first guess' was made, based on the results of fitting the E-dS model. In the E-dS case, the SFH parameter $Q$ did not change and $P$ was slightly increased when density evolution was introduced, so $Q$ was fixed and $P$ allowed to vary. As the best-fitting value of $A_{V_{0}}$ was consistent with local observations, this was also fixed, so allowing only the density evolution parameter, $n$, and $P$ to vary freely. The best-fitting model was $P=3.4, Q=9.0, n=1.1, A_{V_{0}}=0.4$. As can be seen from Fig. \ref{fig:BNSL}, this model is a very good fit to the {\it B} band source counts, so no further variations of the parameters were investigated.

\section{Predicted Counts}
\label{sec:predcounts}

Once the best-fitting sets of parameters in the {\it B} band had been determined ($P=1.5, Q=5.4, n=1.3, A_{V_{0}}=0.4$ for E-dS, $P=3.4, Q=9.0, n=1.1, A_{V_{0}}=0.4$ for $\Lambda$), these sets were used to generate models in other wavebands to determine if the fit in the UV/optical/near-IR wavebands had been improved.

The best-fitting models of RR01 under-predict the faint counts in the UV/optical/near-IR regions. If the new optical depth history correction works, it should increase the goodness of fit in these other bands without tuning the free parameters.

\begin{figure}
 \epsfysize=7.5cm
 \epsfxsize=8.0cm
 \epsffile{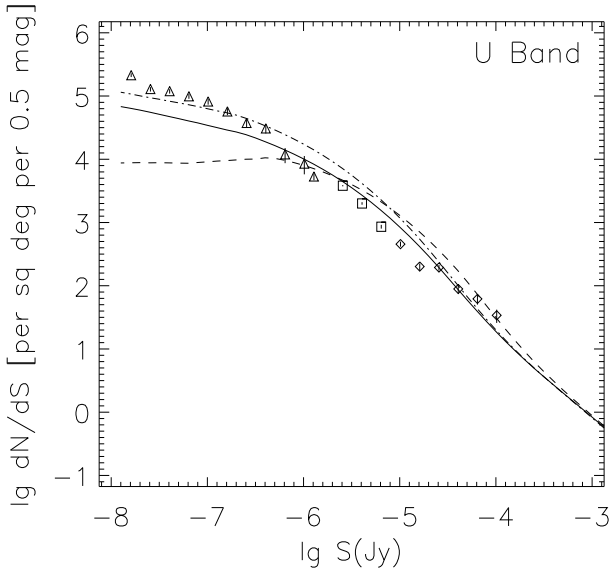}
 \caption{Differential source counts at 0.36-$\umu$m. Data from Koo (1986) [squares], Jones et al. (1991) [diamonds] and Metcalfe et al. (2001) [triangles]. Models are the best-fitting E-dS from RR01, $P = 1.2, Q = 5.4$ [dashed curve], the best-fitting E-dS from this work, $P = 1.5, Q = 5.4, n = 1.3, A_{V_{0}} = 0.4$ [solid curve] and the best-fitting $\Lambda$ from this work, $P = 3.4, Q = 9.0, n = 1.1, A_{V_{0}} = 0.4$ [dotted-dashed curve].}
 \label{fig:UNS}
\end{figure}

\begin{figure}
 \epsfysize=7.5cm
 \epsfxsize=8.0cm
 \epsffile{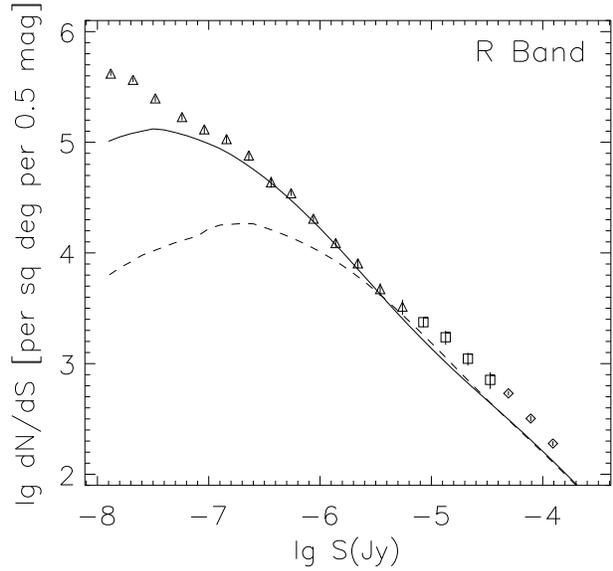}
 \caption{Differential source counts at 0.64-$\umu$m. Data from Jones et al. (1991) [diamonds], Metcalfe et al. (1991, 2001) [squares, triangles]. Models are the best-fitting E-dS from RR01, $P = 1.2, Q = 5.4$ [dashed curve] and the best-fitting E-dS from this work, $P = 1.5, Q = 5.4, n = 1.3, A_{V_{0}} = 0.4$ [solid curve].}
 \label{fig:RNS}
\end{figure}

\begin{figure}
 \epsfysize=7.5cm
 \epsfxsize=8.0cm
 \epsffile{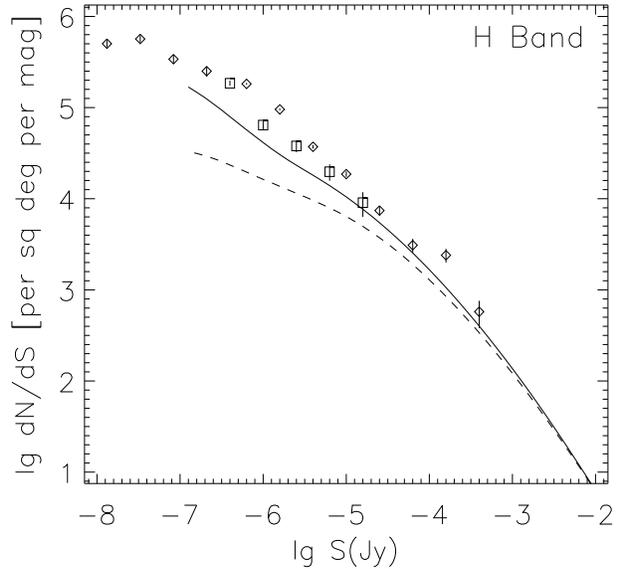}
 \caption{Differential source counts at 1.6-$\umu$m. Data from Metcalfe et al. (in preparation) [diamonds] and Yan et al. (1998) [squares]. Models are the best-fitting E-dS from RR01, $P = 1.2, Q = 5.4$ [dashed curve] and the best-fitting E-dS from this work, $P = 1.5, Q = 5.4, n = 1.3, A_{V_{0}} = 0.4$ [solid curve].}
 \label{fig:HNS}
\end{figure}

\begin{figure}
 \epsfysize=7.5cm
 \epsfxsize=8.0cm
 \epsffile{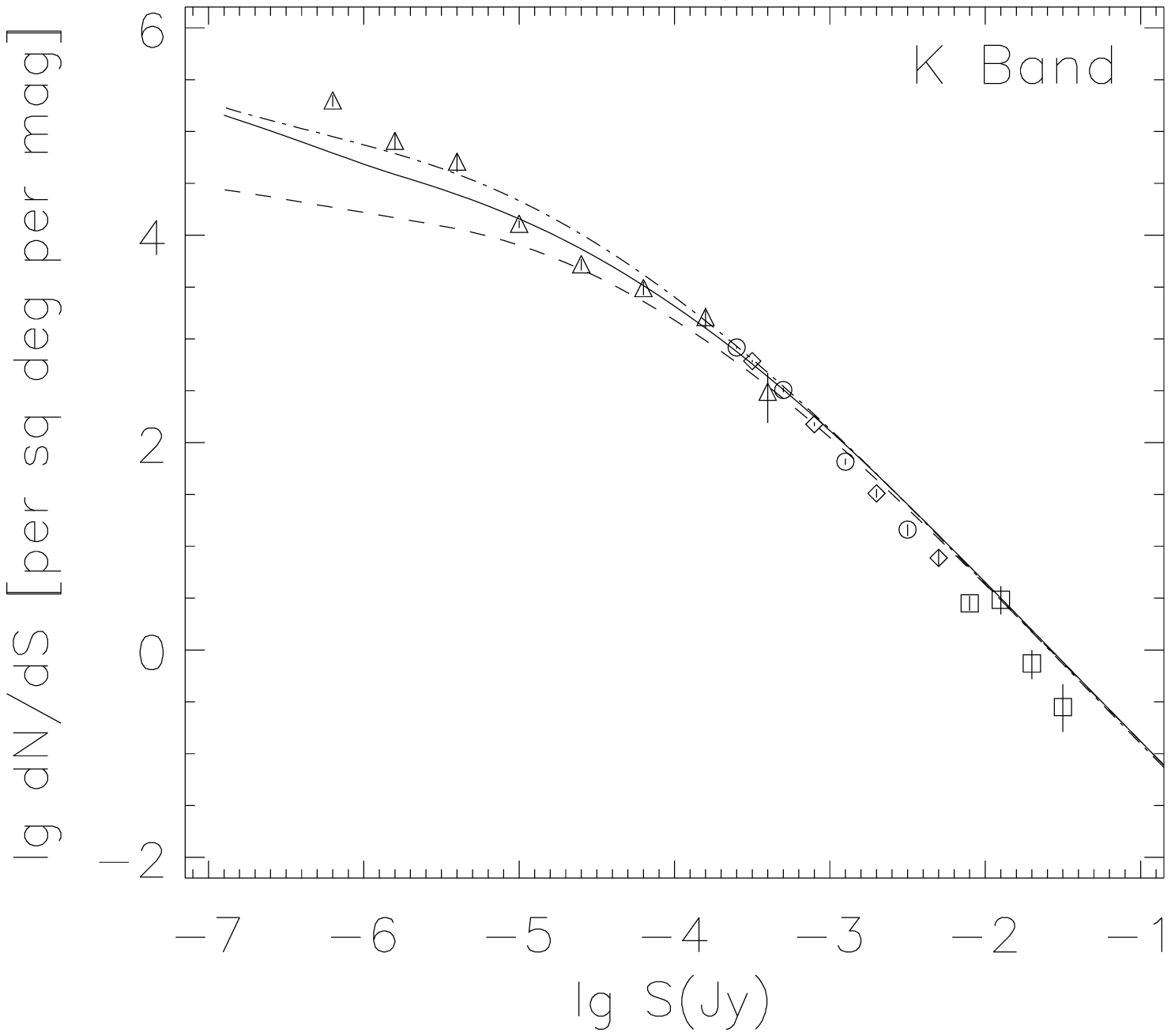}
 \caption{Differential source counts at 2.2-$\umu$m. Data from Mobasher, Ellis \& Sharples (1986) [squares], Gardner et al. (1996) [diamonds], Huang et al. (1997) [crosses] and McCracken et al. (2000) [triangles]. Models are the best-fitting E-dS from RR01, $P = 1.2, Q = 5.4$ [dashed curve], the best-fitting E-dS from this work, $P = 1.5, Q = 5.4, n = 1.3, A_{V_{0}} = 0.4$ [solid curve] and the best-fitting $\Lambda$ from this work, $P = 3.4, Q = 9.0, n = 1.1, A_{V_{0}} = 0.4$ [dotted-dashed curve].}
 \label{fig:KNS}
\end{figure}

The fits to the {\it U} band is shown in Fig. \ref{fig:UNS}. The overall goodness of fit has been improved significantly across the whole flux range for both cosmologies, although the faint ($U > 24$) counts are still under-predicted in the E-dS case. Figs. \ref{fig:RNS}, \ref{fig:HNS} and \ref{fig:KNS} show the fits to the {\it R, H} and {\it K} bands, respectively. In each case the goodness of fit has been substantially improved for both cosmologies, though the very faintest counts tend to be under-predicted in the E-dS case.

\begin{figure}
 \epsfysize=7.5cm
 \epsfxsize=8.0cm
 \epsffile{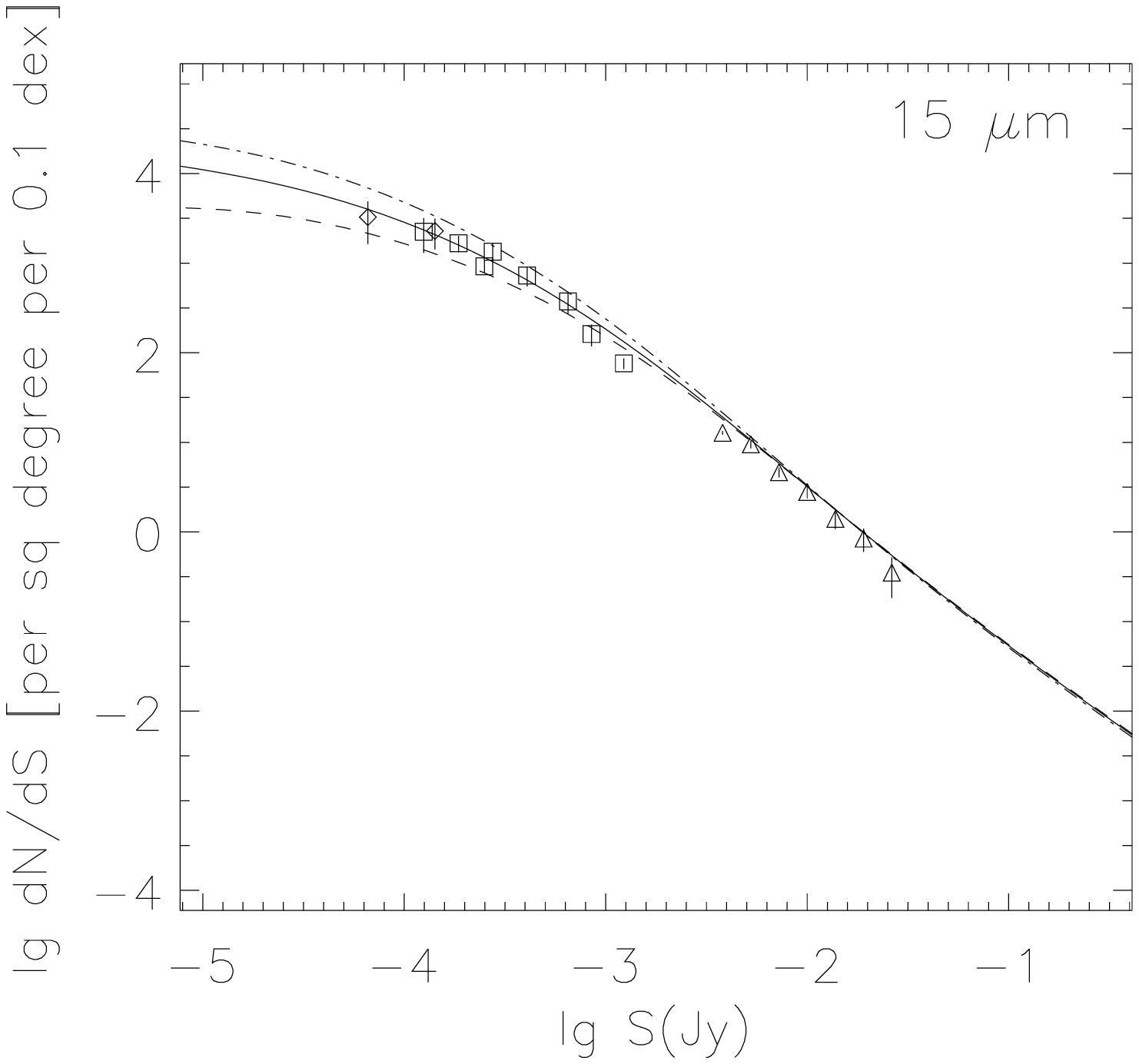}
 \caption{Differential source counts at 15-$\umu$m. Data from Altieri et al. (1999) [diamonds], Elbaz et al. (1999) [squares] and Serjeant et al. (2000) [triangles]. Models are the best-fitting E-dS from RR01, $P = 1.2, Q = 5.4$ [dashed curve], the best-fitting E-dS from this work, $P = 1.5, Q = 5.4, n = 1.3, A_{V_{0}} = 0.4$ [solid curve] and the best-fitting $\Lambda$ from this work, $P = 3.4, Q = 9.0, n = 1.1, A_{V_{0}} = 0.4$ [dotted-dashed curve].}
 \label{fig:15NS}
\end{figure}

\begin{figure}
 \epsfysize=7.5cm
 \epsfxsize=8.0cm
 \epsffile{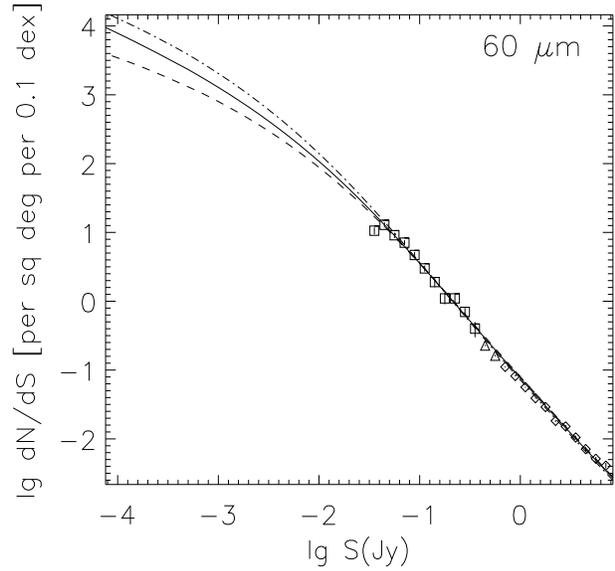}
 \caption{Differential source counts at 60-$\umu$m. Data from Lonsdale et al. (1990) [triangles], Rowan-Robinson et al. (1990) [diamonds] and Gregorich et al. (1995) [squares]. Models are the best-fitting E-dS from RR01, $P = 1.2, Q = 5.4$ [dashed curve], the best-fitting E-dS from this work, $P = 1.5, Q = 5.4, n = 1.3, A_{V_{0}} = 0.4$ [solid curve] and the best-fitting $\Lambda$ from this work, $P = 3.4, Q = 9.0, n = 1.1, A_{V_{0}} = 0.4$ [dotted-dashed curve].}
 \label{fig:60NS}
\end{figure}

\begin{figure}
 \epsfysize=7.5cm
 \epsfxsize=8.0cm
 \epsffile{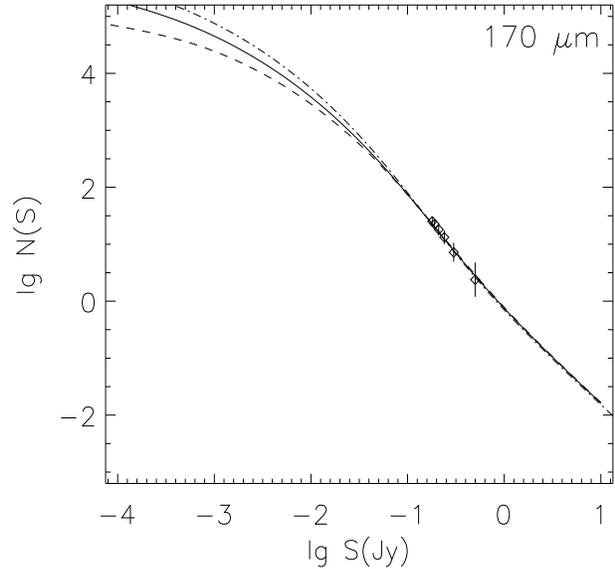}
 \caption{Integral source counts at 170-$\umu$m. Data from Dole et al. (2001). Models are the best-fitting E-dS from RR01, $P = 1.2, Q = 5.4$ [dashed curve], the best-fitting E-dS from this work, $P = 1.5, Q = 5.4, n = 1.3, A_{V_{0}} = 0.4$ [solid curve] and the best-fitting $\Lambda$ from this work, $P = 3.4, Q = 9.0, n = 1.1, A_{V_{0}} = 0.4$ [dotted-dashed curve].}
 \label{fig:170NS}
\end{figure}

\begin{figure}
 \epsfysize=7.5cm
 \epsfxsize=8.0cm
 \epsffile{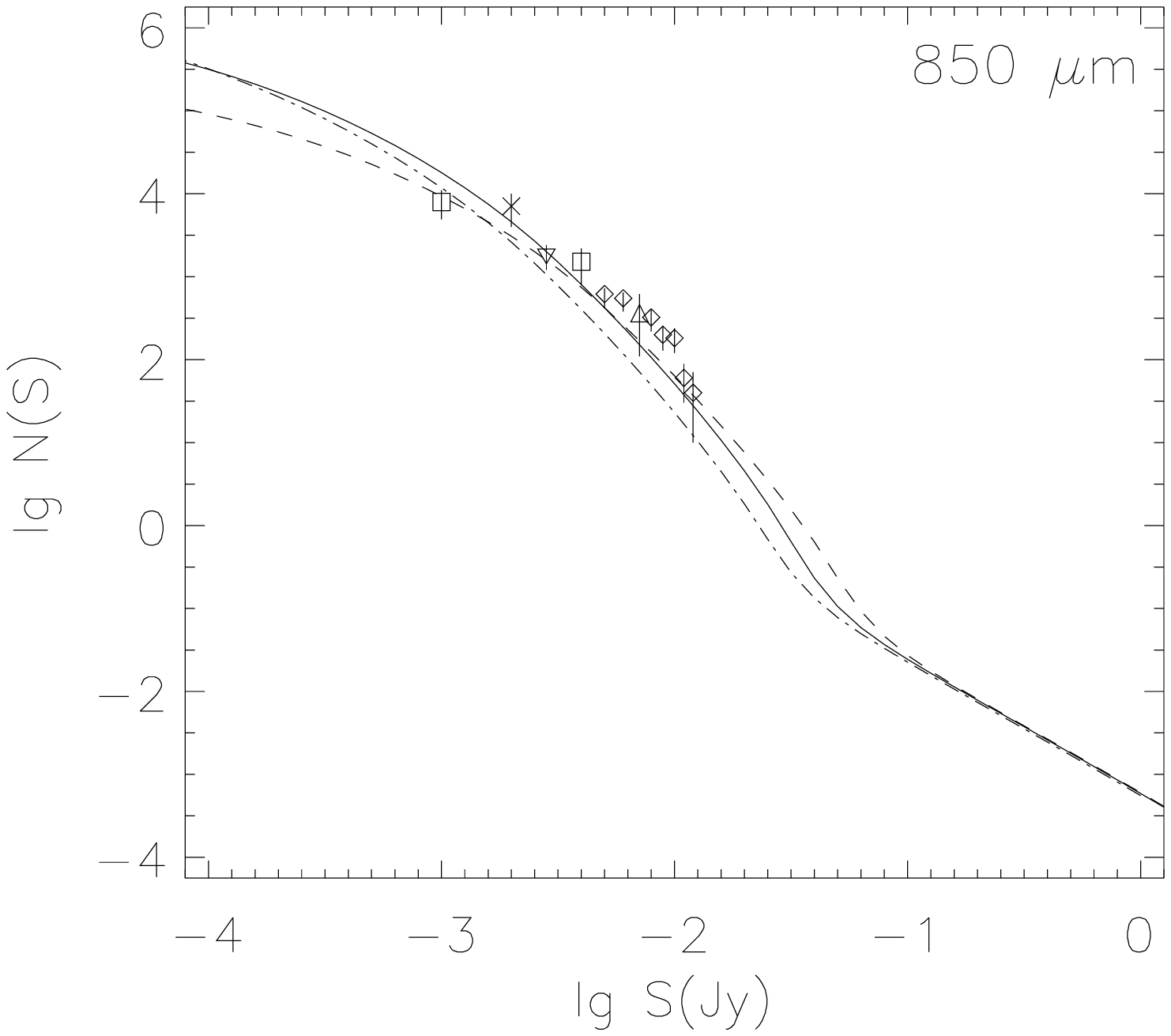}
 \caption{Integral source counts at 850-$\umu$m. Data from Hughes et al. (1998) [cross], Blain et al. (1999a) [squares], Eales et al. (1999) [inverted triangle], Fox (2000) [triangle] and Scott et al. (2002) [diamonds]. Models are the best-fitting E-dS from RR01, $P = 1.2, Q = 5.4$ [dashed curve], the best-fitting E-dS from this work, $P = 1.5, Q = 5.4, n = 1.3, A_{V_{0}} = 0.4$ [solid curve] and the best-fitting $\Lambda$ from this work, $P = 3.4, Q = 9.0, n = 1.1, A_{V_{0}} = 0.4$ [dotted-dashed curve].}
 \label{fig:850NS}
\end{figure}

The $A_{V}$ correction is wavelength-dependent and is negligible for $\lambda > 5\ \umu$m, so this will have no effect on source counts at mid-IR, far-IR and submillimetre wavelengths. However, the values of $P$ and $n$ are different to the preferred models of RR01, which will introduce changes. Predictions at 15, 60, 170 and 850 $\umu$m are shown in Figs. \ref{fig:15NS}, \ref{fig:60NS}, \ref{fig:170NS} and \ref{fig:850NS}, respectively. Across all these wavebands, and for both cosmologies, the best-fitting models of this work predict higher values of the faint source counts than the RR01 models. For 60 and 170 $\umu$m, the available observations are not deep enough to be able to distinguish between the models. At 15 and 850 $\umu$m there is no significant change in the overall goodness of fit.

\section{Discussion}
\label{sec:discussion}


\subsection{Fits to Data}
\label{subsec:fits}

The $P=1.5, Q=5.4, n=1.3, A_{V_{0}}=0.4$ model was selected to provide the best possible fit to the {\it B} band source counts (Fig. \ref{fig:BNS}), whilst not worsening the fit of the redshift distribution (Fig. \ref{fig:Bz26}). As expected, this model provides a significant improvement of fit in this band. It does, however, still slightly under-predict the faintest end ($B > 26$) of the source count slope. Models with a higher value of $A_{V_{0}}$ can provide a better fit to the faint end, but significantly decrease the goodness of fit to the redshift distribution. The $P=3.4, Q=9.0, n=1.1, A_{V_{0}}=0.4$ model in the $\Lambda$ cosmology, however, fits the faint end of the counts well.

Having introduced the optical depth variation correction to make the model more physical, and selected the parameters to fit the {\it B} band, the model is then `translated' to other wavebands. The fits in the optical, UV and near-IR bands ({\it U, R, H, K}; Figs. \ref{fig:UNS}, \ref{fig:RNS}, \ref{fig:HNS}, \ref{fig:KNS}, respectively) are all substantially better for both cosmologies, although there is again an under-prediction in the faint ends ($U > 24, R > 25, H > 21, K > 21$) of the slopes for the E-dS model. 

Quasar counts in the optical ({\it B} band) region are shown in Fig. \ref{fig:BNS}. There are, however, insufficient faint counts to be able to distinguish between the models.

As the parameter set used is derived from optical observations, it is possible that the fit in the mid-IR to submillimetre bands might be worsened. Since the goodness of these fits is the main strength of the RR01 model it is important to check the effects of altering the parameter set in these wavebands. At 60 and 170 $\umu$m (Figs. \ref{fig:60NS} and \ref{fig:170NS}), insufficient faint source counts are available to be able to distinguish between the models. At 15-$\umu$m (Fig. \ref{fig:15NS}) the goodness of fit is improved in the faint ($S < 1$ mJy) region for the E-dS cosmology and essentially unchanged for the $\Lambda$ case. At 850-$\umu$m (Fig. \ref{fig:850NS}) the goodness of fit is decreased in both cosmologies, but the models are still within the limits of the observations.

\begin{figure}
 \epsfysize=7.5cm
 \epsfxsize=8.0cm
 \epsffile{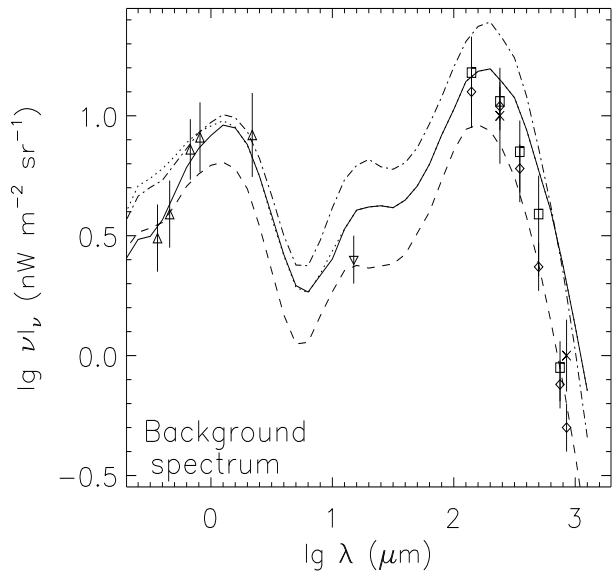}
 \caption{Predicted spectrum of integrated background. Data from Puget et al. (1996) [crosses], Fixsen et al. (1998) [diamonds], Pozzetti et al. (1998) [triangles], Lagache et al. (1999) [squares] and Serjeant et al. (2000) [inverted triangle]. Models are the best-fitting E-dS from RR01, $P = 1.2, Q = 5.4$ [dashed curve], the best-fitting E-dS from this work, $P = 1.5, Q = 5.4, n = 1.3, A_{V_{0}} = 0.4$ [solid curve], $P = 1.5, Q = 5.4, n = 1.3, A_{V_{0}} = 0.0$ [dotted curve] and the best-fitting $\Lambda$ from this work, $P = 3.4, Q = 9.0, n = 1.1, A_{V_{0}} = 0.4$ [dotted-dashed curve]. Note that the fit in the optical is much improved, but the submillimetre counts are over-predicted (see text for discussion of ways to resolve this).}
 \label{fig:spec}
\end{figure}

The best-fitting model was also compared to the available data on the integrated background spectrum (Fig. \ref{fig:spec}). Here, the data is compiled from surveys in the optical and UV (Pozzetti et al. 1998), mid-IR (Serjeant et al. 2000) and far-IR and submillimetre (Puget et al. 1996; Fixsen et al. 1998; Lagache et al. 1999). The new E-dS model [solid curve] shows a better fit in the optical, UV and near-IR regions, but a worse fit in the mid-IR (15-$\umu$m), although this point is derived by summing counts. The goodness of fit in the far-IR (140, 240, 350 and 500 $\umu$m) is approximately the same, but the submillimetre (750 and 850 $\umu$m) background is over-predicted. The dotted curve in this figure shows the comparative $P=1.5, Q=5.4, n=1.3, A_{V_{0}}=0.0$ model to illustrate that it is the optical depth correction, not just the change in $P$ and $n$ that is responsible for the better fit. The $\Lambda$ model is little different to the best-fitting $\Lambda$ model of RR01 - the submillimetre and FIR are over-predicted, but close to the upper limits of the errors.

The over-prediction of faint submillimetre source counts is dominated by the introduction of density evolution. There are certain limitations to the model that might, if corrected for, give a good fit to the optical region with a lower density evolution parameter. Possible candidates for improvement are: 

\begin{enumerate}
 \item Accounting for ellipticals. The count model is based on the 60-$\umu$m luminosity function, which provides a good fit in the infrared. However, this does miss out elliptical galaxies, which have little or no 60-$\umu$m emission. A simplistic way of remedying this would be to multiply the old star component by a factor to account for the missing ellipticals (see \S\ref{subsec:ellip}).
 \item Separately evolving SED components. In the current model, the form of evolution adopted is applied to all the components equally. This simplifies the model and produces reasonable fits to source count data, but is not ideal. Different types of galaxies will evolve differently with redshift, so separating them out would make the model more physical.
 \item Redshift-dependent cirrus SED. The RR01 model does not account for the fact that the temperature of dust, and hence the SED, of cirrus galaxies will evolve with redshift. Including this, perhaps as a look-up table of SEDs at different redshifts, would again make the model more physically realistic.
\end{enumerate}

Introducing one or more of these might allow the density evolution parameter, $n$, to be lowered, and hence the fit to the faint submillimetre counts improved, whilst retaining the goodness of fit in the optical region. The first of these factors is quite simple to correct for and is discussed in \S\ref{subsec:ellip}.


\subsection{Ellipticals}
\label{subsec:ellip}

Initial results from the {\it IRAS} sky survey showed that very few elliptical galaxies had been detected in the FIR (de Jong et al. 1984). Further 

Analysis of the {\it IRAS} sky survey showed that $\sim 35$ per cent of ellipticals are detected at 60-$\umu$m to a flux densities limit of $\sim 300$ mJy (Knapp et al. 1989). As the 60-$\umu$m emission is dominated by emission from cold ($T \sim 20$--$40$ K) dust, this implies that ellipticals typically have low dust densities (e.g. Goudfrooij \& de Jong 1995; Merluzzi 1998). These galaxies will not be accounted for by the 60-$\umu$m luminosity function, which is used by the RR01 model, so some correction is needed. 

Martel, Premadi and Matzner (1998) have used the observed galaxy morphology-density relation (Dressler 1980) and a Monte Carlo technique based on the distribution of dark matter to simulate $\sim$ 40,000 galaxies in the local universe. For an E-dS Universe, they find that ellipticals make up, on average, $\sim 15$ per cent of galaxies. Hence, since only $\sim 35$ per cent of ellipticals are detected at 60-$\umu$m, the RR01 model misses out $\sim 10$ per cent of `cirrus' galaxies.

\begin{figure}
 \epsfysize=7.5cm
 \epsfxsize=8.0cm
 \epsffile{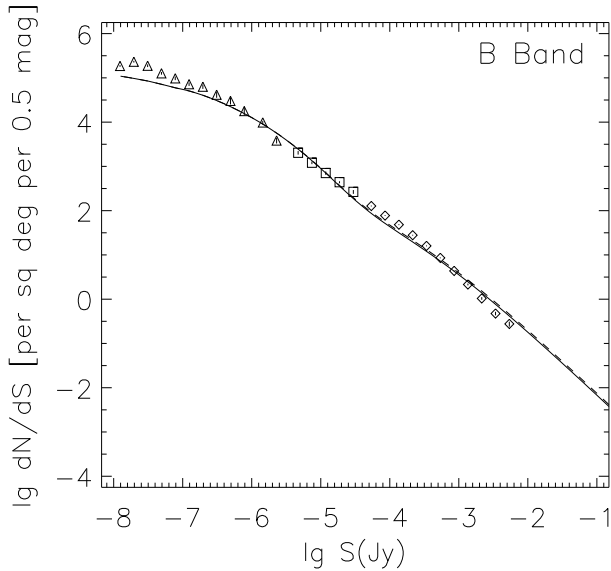}
 \caption{Differential source counts at 0.44-$\umu$m. Data from Jones et al. (1991) [squares], Metcalfe et al. (2001) [triangles] and Yasuda et al. (2001) [diamonds]. Models are the best-fitting from this work, $P = 1.5, Q = 5.4, n = 1.3, A_{V_{0}} = 0.4$ before [solid curve] and after [dashed curve] modification to account for ellipticals. The effect of accounting for ellipticals is negligible.}
 \label{fig:ellip}
\end{figure}

A simple way to correct for this is to multiply the number density of the low-mass cirrus component by a factor of 1.104 at each flux bin for those models that split the cirrus component into low- and high-mass ($\lambda < 20 \umu$m). The result of applying this correction is shown in Fig. \ref{fig:ellip}. The only effect is a very slight increase in the brightest counts, where the low-mass component is dominant. The overall effect of correcting for ellipticals is negligible.

\subsection{Optical Depth}
\label{subsec:opt}

In order to determine how much of the improvement of fit is due to introducing the optical depth correction, and how much is due to the change of $P$ and $n$, Figs. \ref{fig:BNS} and \ref{fig:Bz26} show a model with the same parameter set as the best-fitting E-dS from this work ($P=1.5, Q=5.4, n=1.3$), but with $A_{V_{0}}$ set to zero, so that the optical depth correction is not applied. Fig. \ref{fig:BNS} shows that the alteration in $P$ and $n$ results in a general increase in the number of counts, especially in the fainter regions, whereas the addition of the change in optical depth is responsible for increasing the slope at the faint end of the curves. This shows that it is necessary to include the optical depth correction to achieve this goodness of fit.

The overall goodness of fit to the redshift distribution (Fig. \ref{fig:Bz26}) is little changed between the models. The change in shape is a result of both the optical depth correction and the change in evolutionary parameters.

Observations of local spiral galaxies (e.g. Berlind et. al. 1997; Gonz{\' a}lez et al. 1998) show the visual extinction, $A_{V} \sim 0.1$ -- $0.2$ in the inter-arm regions and $A_{V} \sim 1$ in the arms and centres (Giovanelli et al. 1995; Moriondo, Giovanelli \& Haynes 1998). The best-fitting model of this work has $A_{V_{0}} = 0.4$, which is perfectly reasonable, given known observational results. It is a validation of the optical depth history correction that the value of $A_{V_{0}}$ for the best-fitting model is a physically realistic value purely by fitting to source count data, with no tuning to achieve this result.


\subsection{Star Formation History}
\label{subsec:sfh}

The main strength of the parametrized approach to the SFH, as presented in \S\ref{subsec:evolution} (eqn. \ref{eqn:sfh}), is that it is capable of reproducing most physically realistic, single-population scenarios. This allows a wide range of models to be explored.

\begin{figure}
 \epsfysize=7.5cm
 \epsfxsize=8.0cm
 \epsffile{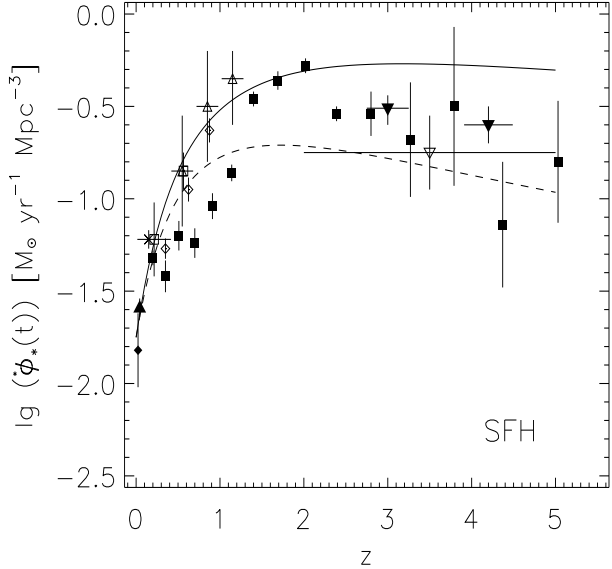}
 \caption{Star formation history in an E-dS cosmology. Models are the best-fitting from RR01, $P = 1.2, Q = 5.4$ [dashed curve], and the best-fitting from this work, $P = 1.5, Q = 5.4, n = 1.3, A_{V_{0}} = 0.4$ [solid curve]. Data points derived from IR (Rowan-Robinson et al. 1997 [open triangles]; Flores et al. 1999 [open diamonds]; Mann et al. 2002 [open squares]), submillimetre (Hughes et al. 1998 [inverted open triangle]), extinction-corrected UV (Gallego et al. 1995 [filled diamond]; Gronwall 1999 [filled triangle]; Steidel et al. 1999 [inverted filled triangles]; Sullivan et al. 2001 [cross]) and re-analysed photometric redshifts of HDF galaxies, corrected for extinction (Rowan-Robinson 2002, in preparation [filled squares]). There is no significant change in the goodness of fit.}
 \label{fig:sfh}
\end{figure}

\begin{figure}
 \epsfysize=7.5cm
 \epsfxsize=8.0cm
 \epsffile{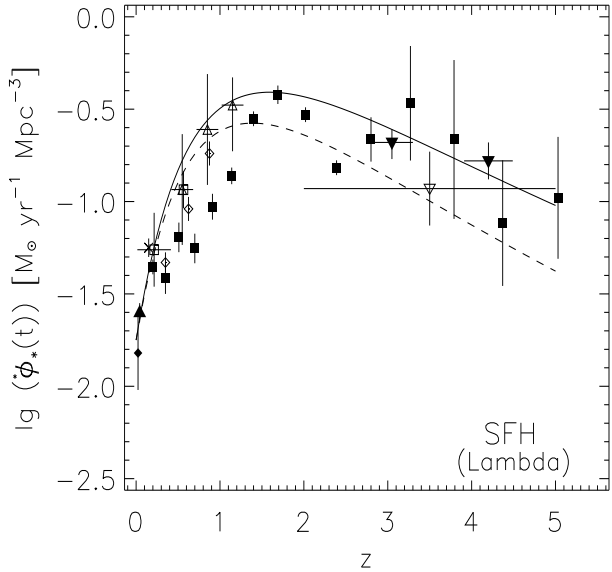}
 \caption{Star formation history in a $\Lambda$ cosmology. Models are the best-fitting from RR01, $P = 3.0, Q = 9.0$ [dashed curve], and the best-fitting from this work, $P = 3.4, Q = 9.0, n = 1.1, A_{V_{0}} = 0.4$ [solid curve]. Data points are derived from infrared data, ultraviolet data with correction for dust and photometric redshifts of HDF galaxies. References are the same as Fig \ref{fig:sfh}.}
 \label{fig:sfh2}
\end{figure}

Once the best-fitting model in the {\it B} band has been determined, the luminosity and density evolutions (eqns. \ref{eqn:sfh} \& \ref{eqn:density}) can be combined to give a star formation history:

\[
 \frac{\dot{\phi_{*}}(t)}{\dot{\phi_{*}}(t_{0})} = \left[\exp\:Q\left(1-\frac{t}{t_{0}}\right)\right]\left(\frac{t}{t_{0}}\right)^{P}(1+z)^{n}.
\]
The predicted SFH for the E-dS cosmology is shown in Fig. \ref{fig:sfh} compared to the $P = 1.2, Q = 5.4$ model of RR01. There is no significant change in the goodness of fit. The SFR at high redshifts ($z > 2$) is probably too high, due to the introduction of density evolution. As with the over-prediction of the faint submillimetre counts, this might be resolved by allowing the components to evolve separately or evolving the SEDs with redshift. For the $\Lambda$ case, shown in Fig. \ref{fig:sfh2}, the goodness of fit is improved in the high-redshift region.

It is important that the form of the star formation history in the model produces a total mass of stars consistent with observed values. Lanzetta, Yahil \& Fern{\' a}ndez-Soto (1996) used optical and UV observations of the Hubble Deep Field to determine the total mass fraction of stars, $\Omega_{*} = 0.0030 \pm 0.0009 h^{-1}$. As this is not corrected for extinction, this is the lower limit on the actual mass of stars. The upper limit is the mass fraction of baryons, which is found by Bania, Rood \& Balser (2002) to be $\Omega_{b} = 0.020_{-0.003}^{+0.007}\ h^{-2}$.

The theoretical value of $\Omega_{*}$ can be calculated from

\[ 
 \Omega_{*} = \frac{\rho_{*}}{\rho_{c}} = 10^{-11.44} h^{-2} t_{0} \dot{\phi_{*}}(t_{0}) \xi,
\]
where 

\[
 \xi = \int_{0}^{1} \frac{\dot{\phi_{*}}(t)}{\dot{\phi_{*}}(t_{0})} d(t/t_{0}).
\]
The SFR in the local universe, $\dot{\phi_{*}}(t_{0})$, is derived from the 60-$\umu$m luminosity function by Rowan-Robinson (2002) using eqn. (7) of Rowan-Robinson et al. (1997), modified to take into account the latest Bruzual \& Charlot galaxy evolution models (Madau et al. 1998), assuming that the fraction of optical-UV light being re-radiated in the IR, $\epsilon = 2/3$:

\[
 \dot{\phi_{*}}(t_{0}) = 10^{-9.66} l_{60},
\]
where the 60-$\umu$m luminosity density, $l_{60} = h \int_{10^{6}}^{10^{14}} \eta (L) dL$. Hence, given $t_{0} = 2/3H_{0}$,

\begin{equation}
 \label{eqn:omeg1}
 \Omega_{*} = 10^{-11.29} \xi l_{60} h^{-3}. 
\end{equation}

For the $P=1.2, Q=5.4$ model of RR01, $l_{60} = 4.38 \times 10^{7}\ h$L$_{\odot}$Mpc$^{-3}$ and $\xi = 5.74$, giving $\Omega_{*} = 0.0013\ h^{-2}$. For the best-fitting model of this work, $l_{60} = 4.58 \times 10^{7}\ h$L$_{\odot}$Mpc$^{-3}$ and $\xi = 12.45$, giving $\Omega_{*} = 0.0029\ h^{-2}$. In this scenario, the $P=1.5, Q=5.4, n=1.3, A_{V_{0}}=0.4$ model is most consistent with observations.

However, the age of the Universe for this calculation has been set to $t_{0} = 6.5$ Gyr, which, although consistent with the cosmology, is not consistent with the observationally-determined age. Using a more realistic value of $t_{0} = 13$ Gyr, eqn. \ref{eqn:omeg1} becomes:
 
\begin{equation}
 \Omega_{*} = 10^{-10.99} \xi l_{60} h^{-2}. 
\end{equation}
This gives $\Omega_{*} = 0.0026\ h^{-1}$ for the best-fitting model of RR01 and $\Omega_{*} = 0.0058\ h^{-1}$ for the best-fitting model of this work. In this case, the RR01 model is more consistent with the observed lower limit value of $\Omega_{*}$. The higher $\Omega_{*}$ in the best-fitting model is due, as with the over-prediction of the submillimetre background, to the inclusion of density evolution.


\subsection{Comparison to semi-analytic model}
\label{subsec:semi}

\begin{figure}
 \epsfysize=7.5cm
 \epsfxsize=8.0cm
 \epsffile{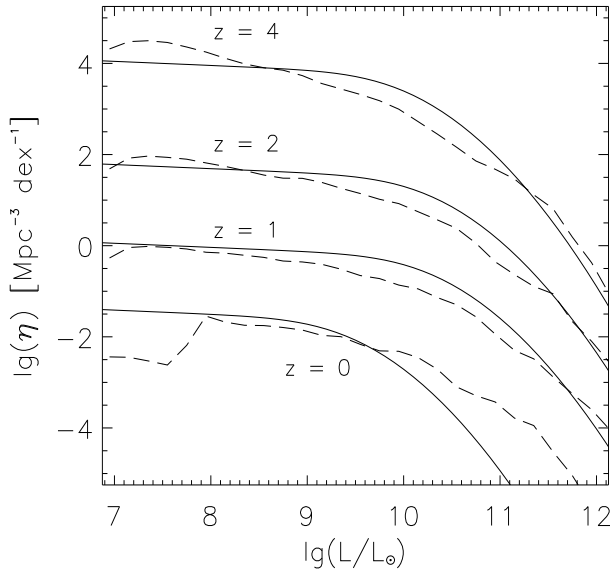}
 \caption{Comparison of RR01 (best-fitting from this work, $P = 1.5, Q = 5.4, n = 1.3, A_{V_{0}} = 0.4$ [solid curve]) and semi-analytic differential luminosity functions (Kaviani et al., 2002 [dashed curve]) at redshifts, z=0.0, 1.0 (offset +1), 2.0 (offset +2.5), 4.0 (offset +4.5). The two models are broadly consistent.}
 \label{fig:lumcomp}
\end{figure}

It is possible to compare the evolution from this parametrized backwards model against a semi-analytic forwards model to determine if they are compatible. Fig. \ref{fig:lumcomp} shows the differential luminosity function at different redshifts. The semi-analytic model used is based on work by Kaviani, Haehnelt \& Kauffmann (2002).

The two luminosity functions are broadly similar, fitting best in the intermediate redshift range $z \sim 1$ -- $2$. The results are broadly consistent, which is interesting, given that they are derived using entirely different methods. A more extensive comparison to the semi-analytic model will be conducted in a future work.

\section{Conclusions}
\label{sec:conclusions}

An attempt to improve the optical count fit of the RR01 IR-defined source count model by introducing density evolution and a variation of optical depth with redshift has been presented.

\begin{enumerate}
 \item By introducing an optical depth correction and tuning the luminosity and density evolution parameters, the goodness of fit to the source counts in the {\it B} band has been significantly increased. This improvement is not due solely to the change in existing parameters, but also substantially affected by the introduction of the optical depth correction and density evolution.
 \item The parameter set derived in the {\it B} band can then be translated to other wavebands, showing significantly improved goodness of fit in the optical, UV and near-IR as well as reasonable fits in the far-IR and submillimetre. 
 \item In the Einstein-de Sitter case, the model still under-predicts the faint end of the source count slope in the optical, UV and near-IR, though to a lesser degree. There is also an over-prediction of very faint submillimetre counts and the high-redshift SFR. These effects are not removed by accounting for ellipticals. It is possible that further corrections to the model could resolve these problems. The $\Lambda$ cosmology does not suffer from these effects, though it does over-predict the FIR region of the background spectrum.
 \item The mean local value of $A_{V}$ of the best-fitting model is 0.4, which is consistent with what we know about the opacity of the local universe.
 \item The form of the luminosity function and the luminosity and density evolutions is broadly consistent with those predicted by a semi-analytic model. 
 \item The predicted form of the SFH provides approximately the same goodness of fit in the E-dS case, and a better fit in the $\Lambda$ case.
 \item It is not possible to to reproduce the multi-wavelength counts using a model without density evolution (the monolithic model).
 \item The model provides a fairly good fit to data from UV to submillimetre. Future work will investigate expanding this to the radio and X-ray regions.
\end{enumerate}


\section*{Acknowledgments}

We are grateful to Duncan Farrah, Matthew Fox, Ali Kaviani, Chris Pearson and the referee for their valuable comments on this manuscript.



\bsp
 
\label{lastpage}

\end{document}